\documentclass[conference]{IEEEtran}

\usepackage{amssymb}
\usepackage{amsmath}

\usepackage{amsthm,theorem}
\usepackage{amssymb}
\usepackage{graphicx,tikz}
\usetikzlibrary{arrows,automata}

\usepackage{array,cite}
\usepackage{subfigure,epsfig}

\title{Collective Sensing-Capacity of Bacteria Populations}
\author{  Arash Einolghozati, Mohsen Sardari, Faramarz Fekri\\
School of Electrical and Computer Engineering\\
Georgia Institute of Technology, Atlanta, GA 30332\\
\texttt{Email:}\{einolghozati, mohsen.sardari, fekri\}@ece.gatech.edu\vspace{-.1in}
\thanks{This material is based upon work supported by the National Science Foundation under Grant No. CNS-111094}
}

\begin{document}
\IEEEoverridecommandlockouts
\maketitle

\begin{abstract}
The design of biological networks using bacteria as the basic elements of the network is initially motivated by a phenomenon called quorum sensing. Through quorum sensing, each bacterium performs sensing the medium and communicating it to others via molecular communication. As a result, bacteria can orchestrate and act collectively and perform tasks impossible otherwise. In this paper, we consider a population of bacteria as a single node in a network. In our version of biological communication networks, such a node would communicate with one another via molecular signals. As a first step toward such networks, this paper focuses on the  study of the transfer of information to the population (i.e., the node) by stimulating it with a concentration of special type of a molecules signal. These molecules trigger a chain of processes inside each bacteria that results in a final output in the form of light or fluorescence. Each stage in the process adds noise to the signal carried to the next stage. Our objective is to measure (compute) the maximum amount of information that we can transfer to the node. This can be viewed as the collective sensing capacity of the node.  The molecular concentration, which carries the information, is the input to the node, which should be estimated by observing the produced light as the output of the node (i.e., the entire population of bacteria forming the node). The molecules are trapped in the bacteria receptors forming complexes inside the bacteria which affect the genes responsible for producing the light. We focus on the noise caused by the random process of trapping molecules at the receptors as well as the variation of outputs of different bacteria in the node. The optimal input distribution to maximize the mutual information between the output of the node, e.g., light, and the applied molecule concentration is derived. Further, the capacity variation with the number of bacteria in the node and the number of receptors per bacteria is obtained. Finally, we investigated the collective sensing capability of the node when a specific form of molecular signaling concentration (which resembles M-ary modulation) is used. The achievable sensing capacity and the corresponding error probabilities were obtained for such practical signaling techniques.
\end{abstract}

\section{Introduction}

The idea of exploiting a colony of bacteria to perform a desired task has been the subject of research in various biological disciplines and new designs have recently attracted the attention of researchers in the information theory and network engineering areas. The basic idea behind such networks is adaptation and engineering of specific types of bacteria that are capable of sensing, computation, actuation, and above all, communication with each other~\cite{Bassler1999}. The existence of a form of communication using molecules that occurs naturally among bacteria has been confirmed~\cite{Bassler1999}. Such a communication enables single cells to gather and process sensory information about their environment and evaluate and react to chemical stimuli. One phenomenon that in particular demonstrates the essential components of communication among cells is ``Quorum Sensing"~\cite{kaplan1985,Bassler1999}. Quorum Sensing can be viewed as a decentralized coordination process which allows bacteria to estimate the density of their population in the environment and regulate their behavior accordingly. To estimate the local population density, bacteria release specific signaling molecules.  These signal molecules will then reach the neighboring bacteria providing them with information about other bacteria in the environment.
 As the local density of bacteria increases, so will the concentration of the molecules in the medium. Bacteria have molecule receptors that can estimate the molecular concentration and thus the bacteria population density. Bacteria use quorum sensing to coordinate energy expensive actions that cannot be carried out by a single bacterium. This process, captures most of the important components of a molecular communication system.

New designs are constantly emerging from manipulation of quorum sensing bacteria's DNA. The encoding, sending, and releasing information using living organisms as carriers of data is studied in~\cite{Palacios2011}. The design of biological clocks using quorum sensing is introduced in~\cite{Danino2010}. A model for forming a network via molecular communication is given in~\cite{Akyildiz2011}. These studies have inspired researchers to investigate the communication among bacteria more carefully and also pay attention to information theoretical aspects of bacteria communication.

 In~\cite{ISIT2011_Arash} we studied a discrete noiseless molecular communication channel inspired by diffusion process in the medium where living organisms dwell in. The information is assumed to be encoded in the \emph{alteration of concentration} of molecules in the medium. 
In~\cite{ITW2011_Arash}, we presented models to account for the reception process of molecules in a single bacterial receiver. 
In~\cite{INFOCOM2011_Arash}, we introduced the Molecular Communication Networking (MCN) paradigm where a population of bacteria (i.e., the primitive agents) that are clustered together form the node of a communication network. Such node is considered to be an independent entity and act as a fairly smart node in the network. The information is transferred from one node to another through diffusion of molecules in the medium. The agents in each node coordinate their actions via quorum sensing process. The agents in a node sense the concentration of molecules at steady-state and responds accordingly. This setup enables us to take advantage of primitive agents in a network that is designed to perform a specific task and transfer information. Motivated by the network framework in~\cite{INFOCOM2011_Arash}, in the present paper, we wish to compute the collective sensing capacity of each such node. We should also mention that there exists another line of research which relies on the assumption that information is encoded in the \emph{timing} of emission of molecules that tries to answer analytical questions concerning the communication in nano-scale through that lens~\cite{eckford,rose2011}.

The output of quorum sensing process can be in various forms, for example, the bacteria can emit light or produce Green Fluorescent Protein (GFP) which can be used to convey information to the outside world. In this paper, our goal is to quantify the maximum amount of information that a population of bacteria can convey, through luminescence or fluorescence, given a constant concentration of molecules as its input. This can then be used to design signaling  schemes for engineering bacteria communication networks such as the one in~\cite{INFOCOM2011_Arash}. We consider a population of bacteria (i.e., a node)  that is stimulated with a slow varying concentration of molecules. The output of the node, in the form of light or GFP, is measured to estimate (i.e., to decode the information) the concentration of molecules at the vicinity of the node. We use a probabilistic model to account for discrepancy in behavior of individual bacteria. We obtain the optimal distribution on input concentration (i.e., the signal) that results in maximum mutual information between the input and output (i.e., the maximum sensing capacity). In addition, we present a signaling technique and obtain the resulting sensing capacity of the node versus the corresponding error rate.

%The response of a strand of bacteria to different levels of concentration of molecules are studied in literature. %In~\cite{muller2008}, a set of differential equations are used to account for the process of producing light, from the trapping of %molecules to the transcription of genes. The average dynamic behavior of bacteria in time and also their steady-state behavior are %studied. The experimental results for the behavior of individual bacteria are studied in[] and discrepancy in behavior of individual %bacteria is observed.

The rest of the paper is organized as follows. In Sec.~\ref{sec:background}, we present the problem setup using the related works in biology and bio-networks. Sec.~\ref{sec:capacity} discusses the analysis of the capacity of concentration sensing and Sec.~\ref{sec:modulation} introduces a practical signal technique that can be used to achieve a certain error rate.

\vspace{-.1in}
\section{Problem Setup}
\label{sec:background}

\begin{figure*}
\vspace{-.2in}
\begin{center}
  \subfigure[A bacteria surrounded by AHL molecules]{
  \includegraphics[width=.4\textwidth]{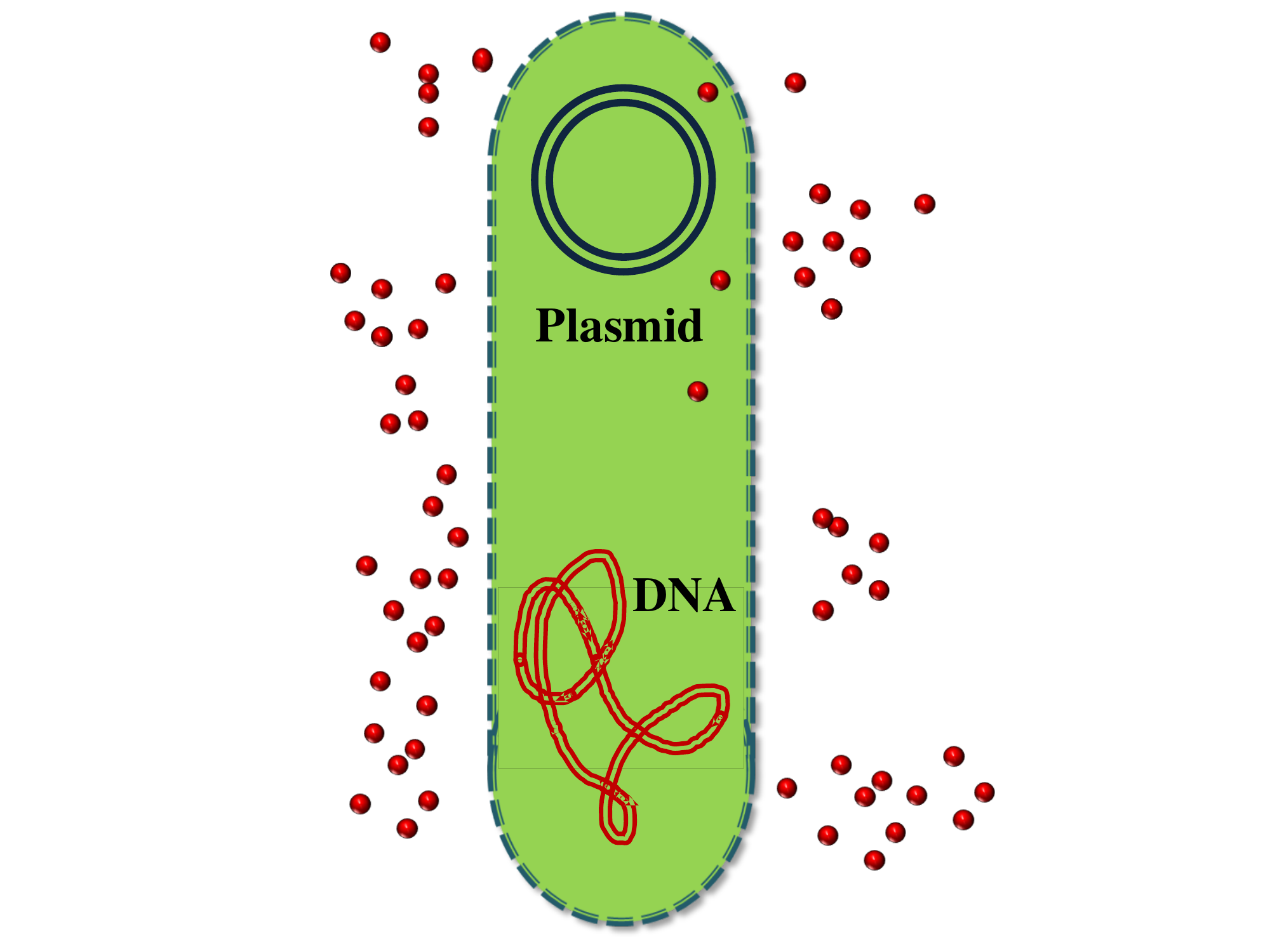}
  \label{fig:bacteria_model}
  }
 % \hspace{-0.2\textwidth}
  \subfigure[The main stages in production of light by bacteria through reception of AHL molecules]{
  \includegraphics[width=.4\textwidth]{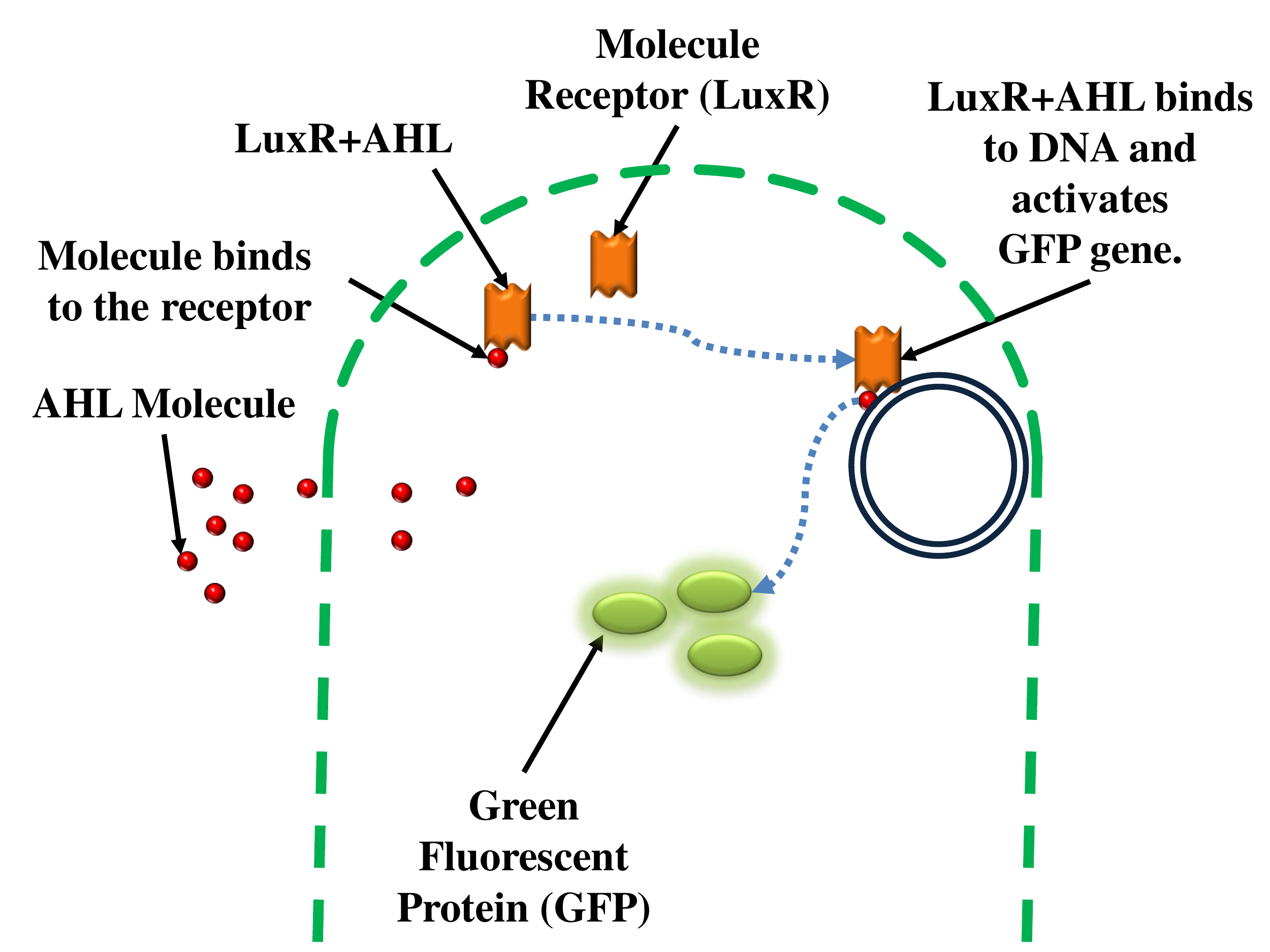}
  \label{fig:bacteria}
  }
\end{center}
\vspace{-.1in}
\caption{Schematic of a bacteria and processes involved in producing GFP}
\label{fig:network_model}
\vspace{-.2in}
\end{figure*}

In our previous work~\cite{INFOCOM2011_Arash}, we introduced a molecular communication network consisting of nodes which contain $n$ bacteria.  The information is conveyed in the network from one node to another through alternation of concentration of Acyl Homoserine-Lactone (AHL) molecules $A$ surrounding the nodes. Note that the random variable $A$ represents the concentration. This concentration is sensed by the bacteria in the node such that each bacterium emits light or GFP with the intensity that depends on $A$. Fig.~\ref{fig:bacteria_model} shows a bacterium used in such a node. The plasmid contains the information (genes) about production of light and can be added to a bacterium that does not naturally emit light.  Each node is a fairly smart unit that can measure the total output (e.g., in the form of light) of the bacteria population residing in the unit and  coordinate their action to alter the concentration of molecular signals (by emitting new molecules to the medium). Note that, in general, a node can sense the concentration of one molecule type and emit to the medium another type of molecular signal (i.e., decode and forward the information to the next node without interference between the input and the output of a node). By this architecture, we form a reliable node out of collection of unreliable bacteria. 

By measuring the output of bacteria (which as mentioned before is in the form of light or GFP), the node is able to estimate the concentration of molecules and hence, decode the information. The number of differentiable levels of input and the error in decoding to the right level limit the amount of information that can be transferred to the node (i.e., sensed by the node). In this setup, one of the fundamental problems is the maximum amount of information that can be transferred to a node and the way it can be achieved. It is important to note that molecular communication networks has very slow dynamics. This is because 1. the molecular diffusion, in the channel connecting two nodes, is very slow, and 2. the chemical processes inside the bacteria are time consuming. Therefore, information delivery from one node to another occurs in a slow pace. Hence, we can assume that molecular signal $A$ varies slowly in time. Therefore, we can assume that the concentration of molecules surrounding a node is constant for sufficient amount of time, allowing the decoding of the information by the node.  Hence, we focus on a receiving node and compute the sensing capacity of a constant concentration of molecules around the node. The setup is consistent with the analysis of the capacity for continuous channels, in which we sample the signal and measure the capacity per sample (rather than per seconds). Finally, note that throughout this paper, we focus on the receiver end and do not study the rate constraints (in bits per seconds) due to the (bandwidth of the) diffusion channel. That is why we referred to the maximum information rate (in bits per sample) as the sensing capacity.

The response of various strands of bacteria to different levels of inter-cellular AHL molecule concentrations has been studied extensively and mathematical models for the chain of chemical reactions it causes  inside the bacterium is provided. 
In~\cite{muller2008}, a model consisting of a chain of linear differential equations is introduced to account for the process of luminescence or fluorescence in response to presence of AHL molecules in the medium. These equations capture the average dynamic behavior of bacteria and also their steady-state behavior. They account for three main stages in the process: 1) binding of AHL molecules to the cell receptors, 2) production of the AHL+LuxR complex and transcription of genes responsible for production of light or GFP and finally, 3)  light emission or GFP production. These three stages are shown in Fig.~\ref{fig:bacteria}. The differential  equation accounting for probability of binding of molecules to the cell receptors is given by~\cite{muller2008}:
\vspace{-.15in}
\begin{equation}
\label{eq:diff}
\dot{p}=-\kappa p+A\gamma \left(1-p\right),
\end{equation}
where $A$ is the concentration of molecules, $\gamma$ is the input gain and $\kappa$ is the dissociation rate of trapped molecules in the cell receptors. Here, $\dot{p}$ is the derivative of $p$ with respect to time. In this model, each cell receptor (i.e., the ligand receptor) is  activated with a probability that depends on the concentration of molecules in the medium surrounding the cell. This probability starts to increase from the moment a constant concentration is applied until it takes its final steady-state value $p_0$, given by
\begin{equation}
\label{eq:steady_state}
p_0=\frac{A\gamma}{A\gamma+\kappa} 
\end{equation}
 Note that $p_0$ increases monotonically with respect to $A$ and approaches $1$ for high concentrations. The process of production of complex molecules and transcription of genes and the process of production of light are modeled similarly. We have
\vspace{-.05in}
\begin{equation}
\left\{
\begin{array}{rcl}
\dot{S}_1 &=& (bp+a)-b_1 S_1\\
\dot{S}_2 &=& a_2 S_1-b_2 S_2
\end{array}
\right.
\label{eq:diff2}
.\end{equation}
where $S_1$ and $S_2$ are two post-transcription messengers and $b_i$ and $a_i$ are some constants~\cite{muller2008}. Note that in the steady state, the value of $S_2$, i.e., the amount of GFP, is a linear function of $p_0$. As explained before, in the analysis of sensing capacity, relations~(\ref{eq:diff}) and~(\ref{eq:diff2}) would be used at steady state.

The probability of binding of molecules maps to the average number of activated receptors. The effect of an activated receptor in the output is additive and total (GFP) output is a linear function of the number of activated receptors in the steady state. The average behavior of the output light is studied in~\cite{muller2008} and the parameters are found by graph fitting. 

For our study, we model the discrepancy in the individual behavior of a population of bacteria in the node. It means that even though the average behavior of bacteria can be formulated with a set of deterministic differential equations, the individual behavior of bacteria features randomness. This randomness is caused by the variation in the various steps of light production process, from molecule binding to gene transcription which changes from one bacterium to another. Such randomness can be accounted for by considering the constants in~\eqref{eq:diff} and~\eqref{eq:diff2} as random variables. For our study, we only consider the noise that is caused by the reception of molecules, as this process is the most closely related one to the input concentration. We intend to study the limits this stage of the light production process imposes on the mutual information between the AHL input and the light output from the entire population of bacteria in a node. We consider two factors that make up the uncertainty of the output of a node. First is the probabilistic nature of the activation of each receptor which enables us to consider each receptor being active as a Bernoulli random variable that is $1$ with probability $p_0$ . The second factor is the variation of $p_0$ throughout the population which is resulted because of the dependency of the input gain to the individual bacterium. As discussed above, the output light of a population  linearly depends on the total number of activated receptors from the entire bacteria population in a node. Hence, the number of activated receptors $Y$ can be considered as the output of a node. In this paper, we study the amount of information that the output $Y$ can give about the input $A$.

\vspace{-.1in}
\section{Sensing Capacity Analysis}
\label{sec:capacity}

As discussed previously, we incorporate the effect of the discrepancy in the behavior of  bacteria into the difference in the molecule reception process. Here, we only consider the noise in the input gain $\gamma$. This model can be generalized to include noise on other parameters in the process as well. We assume that $\gamma$ in~(\ref{eq:steady_state}) has an iid additive noise $\epsilon$ to account for behavior of different  bacteria. The noise $\epsilon$ is caused by uncertainty in performance of receptors of different bacteria. Hence, the entrapment probability can be written as
\begin{equation}
p^*=\frac{A(\gamma+\epsilon)}{A(\gamma+\epsilon)+\kappa}\simeq p_0(1+\frac{\epsilon}{\gamma}-\frac{\epsilon}{\gamma+\frac{\kappa}{A}})
\end{equation}
where $p_0$ is defined in~(\ref{eq:steady_state}) and $\epsilon$ is a zero mean noise with small variance $\sigma^2$ such that we can ignore the second and higher orders of $(\frac{\epsilon}{\gamma})$. We assume this noise affects all the receptors of a bacterium in the same way but it is iid for different bacteria in a population. With some manipulations, we can see
\vspace{-.03in}
\begin{equation}
\label{eq:noise_p}
 p^*=p_0+\frac{\epsilon}{\gamma}p_0(1-p_0)
\end{equation}
 where the second term is an additive noise in $p^*$ whose variance depends on $p_0$. 
We denote by $Y_{ij}$ the random variable representing the status of $j^{th}$ receptor of $i^{th}$ bacterium. In other words, $Y_{ij}$ is a Bernoulli random variable that is $1$ with probability $p_i=p_0+e_i$. Here, $e_i$ is a zero-mean noise with variance $\sigma_0^2 p_0^2 (1-p_0)^2$ where $\sigma_0^2=\frac{\sigma^2}{\gamma^2}$ is the normalized power of noise. We denote by $n$ the number of bacteria in the population residing in a node and by $N$ the number of receptors per each bacterium. Hence, the output will be
\begin{equation}
\label{eq:output}
Y=\sum_{i=1}^n Y_i =\sum_{i=1}^n \sum_{j=1}^N Y_{ij}
\end{equation}
where $Y_i$, (i.e., $Y_i=\sum_{j=1}^N Y_{ij}$) is the output of bacterium $i$. Using the conditional expectation, we have
\begin{equation}
\label{eq:exoected_value}
E(Y_i)=E(E(Y_i|p_i)))=E(Np_i)=Np_0
\end{equation}
where the last equality comes from the fact that the noise has zero mean. Hence, for the output defined in~(\ref{eq:output}), we have E(Y)=$n N p_0$. By using conditional variance, we have
\vspace{-.05in}
\begin{align}
\label{eq:variance}\nonumber
Var(Y_i) &=E(Var(Y_i|p_i))+Var(E(Y_i|p_i))\\
	\nonumber&=E(N p_i(1-p_i))+Var(Np_i)\\
&=N p_0(1 -p_0)+(N^2-N) \sigma^2 p_0^2 (1-p_0)^2
\end{align}
The first term in~(\ref{eq:variance}) is due to general uncertainty in a Binomial output and the second one is due to noise in the parameter $p_i$. By independence assumption between the output of different bacteria, the variance of the output is equal to
\begin{equation}
\label{eq:variance_output}
 \sigma_Y^2=nN p_0(1 -p_0)+n(N^2-N) \sigma^2 p_0^2 (1-p_0)^2. 
\end{equation}
Since the number of receptors $N$ per bacterium is usually large enough, the second term is dominating. Hence, we can approximate the variance by $n N^2 \sigma^2 p_0^2 (1-p_0)^2$.

Note that $Y$ can take $nN+1$ different levels which results in a trivial upper bound of $\log_2 (nN+1)$ for the mutual information between $A$ and $Y$. In order to make the analysis of the capacity tractable, we approximate the output in~(\ref{eq:output}) with a Normal random variable. Since the number of receptors $N$ is large, we can use the Central Limit Theorem and approximate $Y_i$ by ${\bf \mathcal{N}} (Np_0,\sigma_{Y_i}^2)$ where $\sigma_{Y_i}^2$ is given in~(\ref{eq:variance}). Hence, the output $Y$ will be the sum of $n$ Normal variables given by
\begin{equation}
\label{eq:normal_output} 
Y={\bf \mathcal{N}}(nNp_0,\sigma_Y^2)=nNp_0+\epsilon_Y
\end{equation}
where the noise $\epsilon_Y\sim{\bf \mathcal{N}}(0,\sigma_Y^2)$. The first term in~(\ref{eq:normal_output}) can be considered as the signal to be decoded by the node and the second term is an additive Gaussian noise which has a signal-dependent variance. In order to calculate the sensing capacity, we should obtain the optimized distribution of $p_0$ which maximizes $I(p_0;Y)$ the mutual information between the input and output, which in turn gives the optimized distribution for $A$ through~(\ref{eq:steady_state}).
%%%%%%%%%%%%%%%%%%%%%%%%%
\begin{figure}
\includegraphics[width=.48\textwidth]{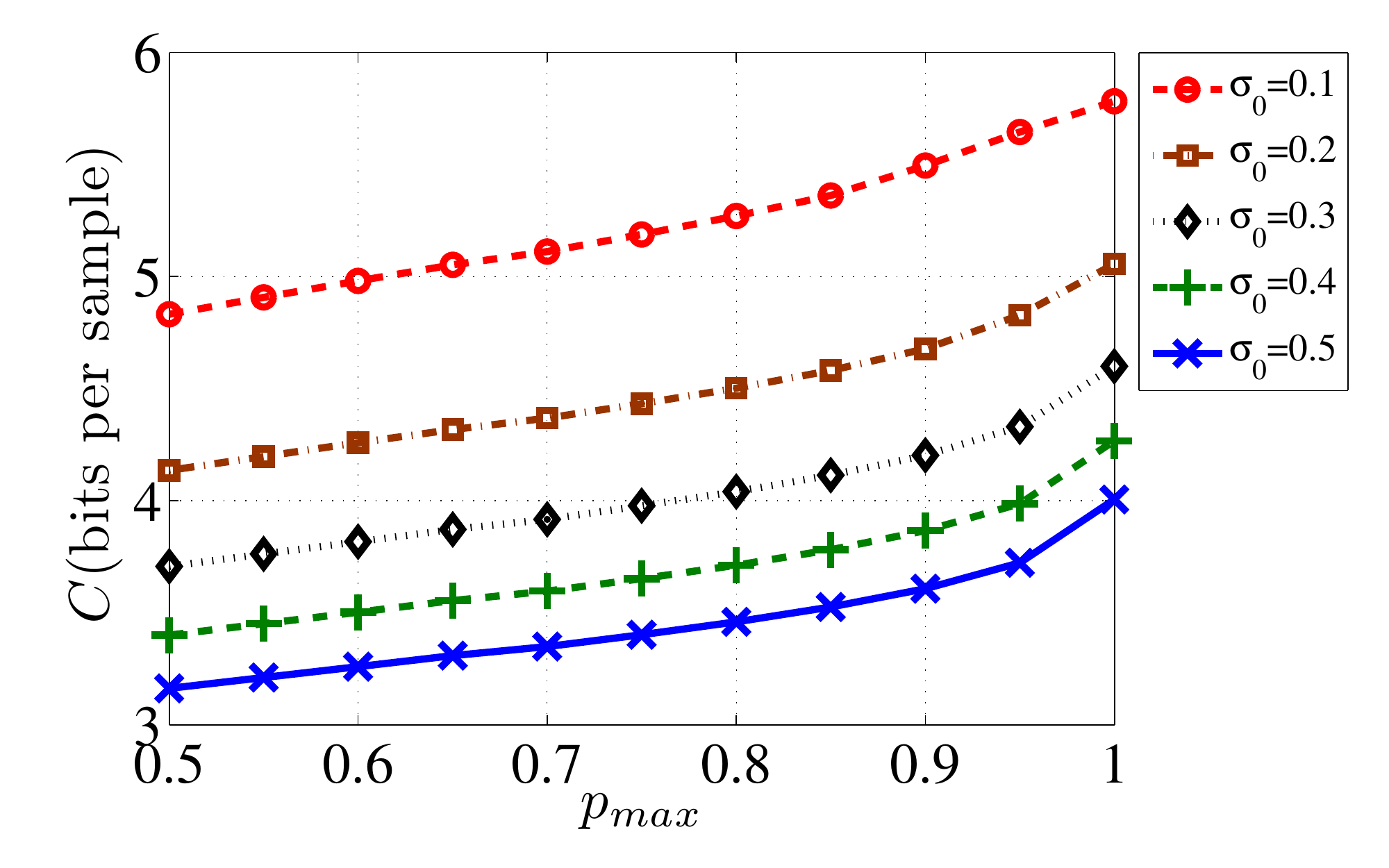}
\vspace{-.17in}
\caption{Capacity (bits per sample) versus maximum trapping probability $p_{max}$ for different values of noise power$\sigma_0^2$ }
\vspace{-.17in}
\label{fig:capacity_noise}
\end{figure}

\begin{figure}
\centering
\includegraphics[width=.48\textwidth]{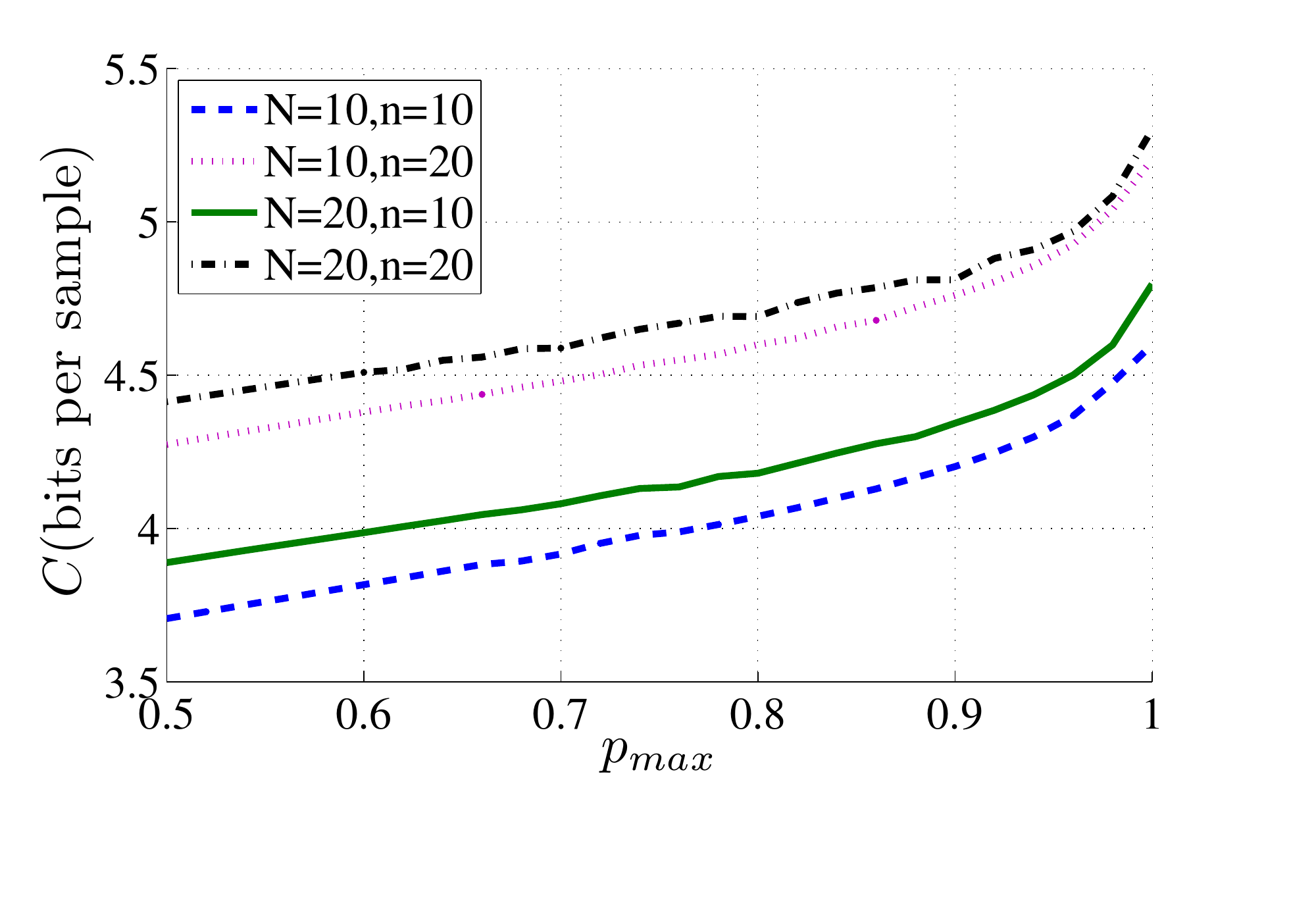}
\vspace{-.4in}
\caption{Capacity (bits per sample) versus maximum trapping probability $p_{max}$, plotted for different number of bacteria $n$, and the number of receptors per bacterium $N$ }
\vspace{-.3in}
\label{fig:capacity_number}
\end{figure}
%%%%%%%%%%%%%%%%%%%%%%%%%

 To proceed, we observe that, in practice, $A$ cannot take any values and the maximum achievable concentration is $A_{max}$ which maps to probability $p_{max}$ via~(\ref{eq:steady_state}). This maximum probability is an indicator of the maximum power used by the transmitter to emit the molecules into the medium. By using more power, the transmitter can increase the maximum concentration of molecules at the vicinity of the receiver node and increase $p_{max}$. We obtain the optimized distribution for $p_0$ over the interval $ [0 \;\; p_{max}]$ and calculate the capacity based on $p_{max}$. We assume $p_{max}$ to be greater than $\frac{1}{2}$.

The structure of noise in~(\ref{eq:normal_output}) is complicated since the noise variance depends on the signal itself. Hence, we use the numerical method of Blahut-Arimoto algorithm (BA) in order to obtain the optimal input distribution and its corresponding capacity. Note that due to above approximations, it will give us the approximated sensing capacity. As we can see in~(\ref{eq:variance_output}), the noise power is at its maximum at $p_0=\frac{1}{2}$ and goes to zero when $p_0$ approaches to either zero or one. Hence, we expect that the distribution of $p_0$ should take values close to $0$ and $p_{max}$ with a higher probability. The results from the algorithm confirms this fact and the distribution has local maximums at $0$ and $p_{max}$.  

Results for the capacity (in bits per sample) with respect to $p_{max}$ for different values of $\sigma_0^2$ is shown in Fig.~\ref{fig:capacity_noise}. In this setup, we have chosen $N=n=30$. As we observe from the plot, the capacity increases when we either increase $p_{max}$ or decrease $\sigma_0^2$. Note that the maximum achievable capacity is limited even if the transmitter used infinite power to make $p_{max}=1$. Here, we have assumed $\sigma_0^2$ to be nonzero but even in the case that $\sigma_0^2=0$, the capacity is limited. This is because of the first term in the variance of noise in~(\ref{eq:variance_output}) which does not depend on $\sigma_0$ and is caused by binomial nature of the reception process. Note that in practice, $N$ and $n$ are very large. However due to the exponential growth of the simulation time with respect to $N$ or $n$, we only computed the capacity for small values of $N$ and $n$.

Fig.~\ref{fig:capacity_number} shows the capacity versus $p_{max}$ for different values of $n$ the number of bacteria in the population, and $N$ the number of receptors per bacterium. Here, $\sigma_0^2$ is considered to be $0.1$ during the simulation. As we observe from the plot, by increasing $n$ or $N$, the capacity increases but the effect of $n$ is more significant. The reason for that can be explained by~(\ref{eq:variance_output}) and~(\ref{eq:normal_output}). The input signal increases linearly with both $n$ and $N$ but the variance of the noise increases quadratically with $N$ and linearly with $n$.

\vspace{-.05in}
\section{Achievable Rates under Practical Signaling (Modulation)}
\label{sec:modulation}

In the previous section, the capacity was obtained with the assumption that $A$ (and hence $p_0$) is a continuous variable between $0$ and $A_{max}$. In other words, it was assumed that any concentration of molecules can be achieved by the transmitter at the vicinity of the node. In practice, a discrete number of levels of concentration of molecules can be achieved which results in discrete levels of $p_0$.

In this section, we consider a specific signaling method (i.e., modulation) and study the modulations of input signal that achieve specific error rates and compute its achievable information rate. The range of the input is determined by $p_{max}$ at the receptors. The number of levels of concentration and different choices for those levels  introduce different ways of modulation of information. In essence, given $k$, the number of input signal levels, we have to find the best $k$ levels to choose from the interval $[0\quad p_{max}]$ and their corresponding weights in order to minimize the error of transmission.
Since the variance of the error in the output varies for different inputs, the ideal analysis as described above will be cumbersome and actually not practical. This is because of the complexity that it enforces on the transmitter, which is assumed to be simple. Here, we use a rather simple but effective modulation of the input signal in which the $k$ levels are chosen uniformly from the interval $[0\quad p_{max}]$ and symbols are assumed to be equi-probable. The probability of error can be reduced by increasing the transmission power which in turn increases $p_{max}$. Again we assume $p_{max}$ to be always greater than $\frac{1}{2}$.

The most trivial modulation corresponds to $k=2$ in which symbols $p_0=0$ and $p_0=p_{max}$ are transmitted with equal probability. Ideally if the transmitter is able to make $p_{max}=1$, we can achieve the data rate of $1$ with error probability of zero. This is because the symbols $0$ and $1$ do not make any noise in the receiver (see~(\ref{eq:noise_p})). Hence, zero error can be achieved in a binary transmitter with power approaching to infinity.

\begin{figure}
\centering
\vspace{-.1in}
\includegraphics[width=.48\textwidth]{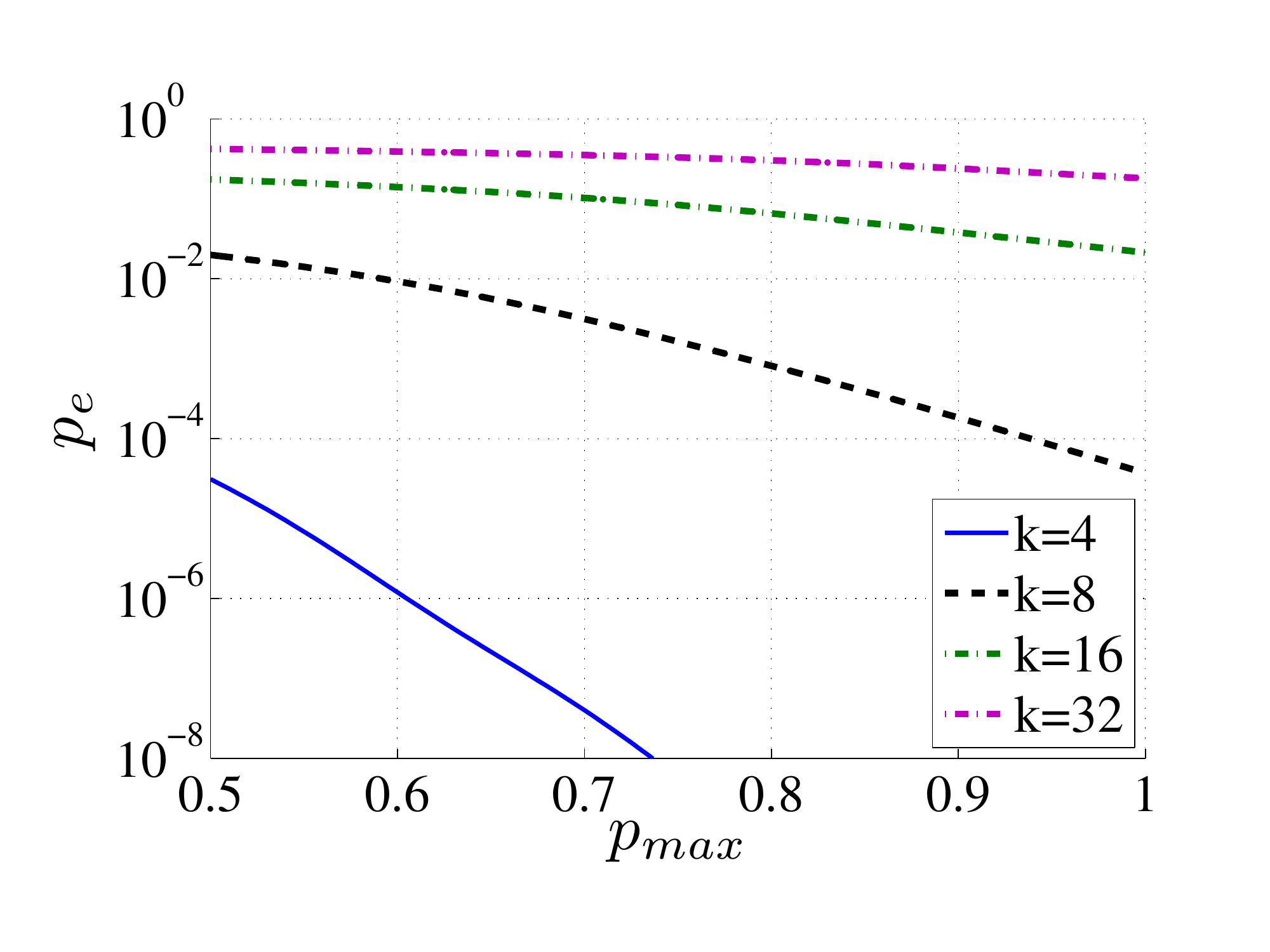}
\vspace{-.3in}
\caption{Probability of error with respect to maximum trapping probability for different signaling techniques }
\vspace{-.2in}
\label{fig:error_probability}
\end{figure}

We consider a hard decision scenario in decoding the symbols. In other words, the receiver chooses the closest symbol to the received one. We obtain the error probabilities $p_e$ versus $p_{max}$ as shown in Fig.~\ref{fig:error_probability} for different modulations. The results are given for $n=N=30$ and $\sigma_0^2=0.1$. The probability of error for $k=2$ is practically zero even for finite power. We can conclude from Fig.~\ref{fig:error_probability}, that the communication is not reliable as $k$ increases. For example, for the case of $k=32$, the probability of error cannot become less than $0.2$. However, by increasing $N$ and $n$, for any $k$, the scheme becomes a practical modulation. The experiment is not plotted due to page limit.
Note that $p_e$ decreases as $p_{max}$ increases but it cannot be made arbitrarily small for a modulation with large $k$, unless $n$ and $N$ are large. 

%\vspace{-.1in}
\section{Conclusion}

In this paper, we studied the transfer of information to a population of bacteria by stimulating them with a concentration of molecules. The  output can be in the form of light or GFP and contains a noise from the process of trapping of the molecules and the variation in bacteria responses. the power can be accounted for by the maximum achieved concentration at the receiver. The optimal input distribution and the corresponding capacity were obtained based on the received power. Finally, we introduced practical modulations that achieve a specific data rate and error probability.

%\vspace{-.1in}
\bibliographystyle{IEEEtran}
\bibliography{ISIT2012}
\end{document}